\documentclass[journal]{IEEEtran}
 
\usepackage{amsmath}
\usepackage{amssymb}
\usepackage{xcolor}
\usepackage{graphicx}
\usepackage[font=small]{caption}
\usepackage[subrefformat=parens]{subcaption}
\usepackage{amsthm}
\usepackage{enumitem}
\usepackage{float}
\usepackage{bm}
\usepackage{enumitem}
\usepackage[noend]{algorithmic}
\usepackage{algorithm}
\usepackage{multirow} 
\usepackage{soul} 
\usepackage{cite}
\usepackage{mathtools}

\newtheorem{theorem}{Theorem}

\newtheorem{example}{Example}
\theoremstyle{definition}
\newtheorem{remark}{Remark}
\newtheorem{definition}{Definition}

\usepackage{tikz}
\usepackage{pgfplots}
\usetikzlibrary{plotmarks}
\usetikzlibrary{matrix,positioning,shapes,arrows,decorations,calc,fit}
\usetikzlibrary{decorations.pathreplacing}
\usetikzlibrary{spy,backgrounds}
\usetikzlibrary{external}
\usetikzlibrary{patterns}
\usetikzlibrary{shapes,arrows}
\usetikzlibrary{shadows.blur}
\usetikzlibrary{shapes.symbols}
\usetikzlibrary{
	pgfplots.groupplots,
	matrix
}

\newcommand{\circC}{\mathbin{\text{\textcircled{c}}}}

\newcommand{\blue}[1]{{\textcolor[rgb]{0,0,1}{#1}}}

\IEEEaftertitletext{\vspace{-3\baselineskip}}

\begin{document}
 
 \title{Optimal Linear MAP Decoding of Convolutional Codes}
 
\author{Yonghui Li,~ Chentao Yue, and Branka Vucetic 

 School of Electrical and Compter Engineering, the University of Sydney, Sydney, NSW 2006 
}

\maketitle

\begin{abstract}

In this paper, we propose a linear representation of BCJR maximum a posteriori probability (MAP) decoding of a rate 1/2 convolutional code (CC), referred to as the linear MAP decoding (LMAP). We discover that the MAP forward and backward decoding can be implemented by the corresponding dual soft input and soft output (SISO) encoders using shift registers. The bidrectional MAP decoding output can be obtained by combining the contents of respective forward and backward dual encoders.  Represented using simple shift-registers, LMAP decoder maps naturally to hardware registers and thus can be easily implemented. Simulation results demonstrate that the LMAP decoding achieves the same performance as the BCJR MAP decoding, but has a significantly reduced decoding delay.  For the block length 64, the CC of the memory length 14 with LMAP decoding surpasses the random coding union (RCU) bound by approximately 0.5 dB at a BLER of $10^{-3}$, and closely approaches both the normal approximation (NA) and meta-converse (MC) bounds.

\end{abstract}

\vspace{-1em}
\section{Introduction}

Convolutional codes (CCs) have been widely used in modern communications systems. There are two commonly used convolutional decoding algorithms, the Viterbi algorithm (VA) \cite{Viterbi} and the BCJR maximum a posteriori probability (MAP) algorithm \cite{BCJR}.  Their key limitations are their high computing complexity. In addition, they require frequent and complex cache and memory access due to metric table and trellis lookups, which significantly contribute to delays and complexity in decoding implementations. Decoding CCs with constraint lengths $\geq 10$ is challenging when using the VA or the MAP algorithms.  Some reduced complexity decoders, such as serial and parallel list Viterbi decoders \cite{LVD1,LVD2,LVD3}, CRC-aided convolutional codes, and Sequential decoders \cite{Fano,stack1,stack2,VBT}, can reduce the decoding complexity, but they are suboptimal and cannot approach the performance of MAP decoder. Also the decoding effort and delay are random for sequential decoders as some codewords require lengthy searches and substantial computations, occasionally surpassing the computational demands of the Viterbi algorithm. This can result in buffer overflows or necessitate large storage capacities. 

In our prior work \cite{ETT1,ETT2}, we have discovered that the MAP forward and backward decoding process for a rate-1 CC can be represented by a respective encoder implemented using shift registers, which we refer to as the dual encoder of forward and backward decoding, respectively. If we apply the soft symbol estimates (SSE) of convolutional coded symbols as the dual encoder input, then its output produces the SSE of information symbols of MAP decoding. We have shown that a decoder using a dual encoder structure is equivalent to the original BCJR MAP decoding, and can significantly reduce the decoder complexity from $O(N^2)$ in the original MAP decoding to $O(\log(N))$.  

In this paper, building on our prior work \cite{ETT1,ETT2}, we develop a linear soft-input soft-output (SISO) MAP decoding for a rate-1/2 CC. To gain insights, we first investigate the MAP forward decoding process of 4-state codes. Similar to the decoding of rate-1 codes \cite{ETT1}\cite{ETT2}, we discovered that the MAP forward decoding of 4-states codes can be implemented by a dual encoder using shift registers.  Then, we generalize the  dual encoder structure to the general rate-1/2 CC and verify the results with different codes. Dual encoder for forward decoding consists of two constituent dual encoders, $\mathrm{DF}_1$ and $\mathrm{DF}_2$. $\mathrm{DF}_1$ and $\mathrm{DF}_2$ compute all the dual encoder memories in different orders using different shift register structures where the inputs of these two dual encoders are the SSE of the received coded symbols. That is, the order of the dual encoder memories in $\mathrm{DF}_1$ and $\mathrm{DF}_2$ are different. We derive the explicit order of memories for each dual encoder. Similarly, we derive the two constituent dual encoders for backward decoding $\mathrm{DB}_1$ and $\mathrm{DB}_2$. The bidirectional MAP decoding output is obtained by linearly combining the shift register contents of $\mathrm{DF}_1$ and $\mathrm{DB}_1$, and that of $\mathrm{DF}_2$ and $\mathrm{DB}_2$. 

The dual encoders can be calculated offline. The LMAP decoding can be implemented using shift register structure, resulting in fast decoding. Matlab simulations demonstrate that the LMAP decoding  achieves the same performance as BCJR MAP decoding, but greatly reduces the decoding time by decreasing the delay and complexity in decoding computation and caching access. It enables a computationally efficient decoding algorithm and unlocks the decoding of very large-state convolutional codes (VLSC), with capacity approaching performance and low complexity.  It shows that for the short block length 64, the VLSC of memory length $m=14$ with LMAP decoding outperforms the RCU bound and closely approaches both the NA and MC bounds. This demonstrates the  near-optimal performance of the code and LMAP decoding.

\vspace{-0.7em}
\section{Linear Representation of MAP Decoding for Rate-1 Convolutional Codes}
\vspace{-0.4em}

In this section, we present the key results of linear representation of MAP decoding for rate-1 CCs \cite{ETT1,ETT2}.  

Let $\mathbf{b}=(b_1,b_2,\ldots,b_L)$ and $\mathbf{c}=(c_1,c_2,\ldots,c_L)$ be a binary information sequence and its codeword, generated by a rate-1 generator polynomial $g(x)$, and $\mathbf{x_c}=(x_{c_1},x_{c_2},\ldots,x_{c_L})$ be the BPSK modulated sequence of $\mathbf{c}$.  Let $\mathbf{y}=(y_1,y_2,\ldots,y_L)$ denote the received sequence. Let  $\hat{x}_{c_k}$ and $\hat{x}_{b_k}$  denote the soft symbol estimate (SSE) of received coded symbol ${x_k}$ and information symbol ${b_k}$ , respectively. We assume that 0 and 1 are modulated into symbol 1 and -1. Then the SSEs $\hat{x}_{c_k}$ and $\hat{x}_{b_k}$,  can be calculated as
\begin{equation} \small
    \hat{x}_{c_k}=p_{c_k}(0)-p_{c_k}(1) \ \ \text{and} \ \ \hat{x}_{b_k}=p_{b_k}(0)-p_{b_k}(1),
\end{equation}
where  $p_{c_k}(l)=p(c_k=l|y_k)$ and $p_{b_k}(w)=p(b_k=w|\mathbf{y})$.  

\begin{definition} 
For a polynomial $a_f(x)$ of degree $n$, we define its \textit{minimum complementary polynomial} (MCP) as the polynomial of the smallest degree $z_f(x)$, i.e.,
\begin{equation} \label{eq7} \small
a_f(x)z_f(x)=x^{n+l}+1.
\end{equation}
It can be shown that $z_f(x)$ always exists.
\end{definition}
\begin{theorem}[Linear Representation of MAP Forward Decoding \cite{ETT1}] For a rate-1 CC, generated by a  $g_f(x)=\frac{a_f(x)}{q_f(x)}$, let $z_f(x)$ be the degree-$l$ MCP of $a_f(x)$. The log-domain SISO forward decoding of the code can be simply implemented by its dual encoder with the generator polynomial of
    \begin{eqnarray} \label{eq9}   \small
    d_{f}(x) &=&
    \frac{q_f(x)z_f(x)}{a_f(x)z_f(x)}=\frac{q_f(x)z_f(x)}{x^{n+l}+1}
    \end{eqnarray}
where the decoder inputs and outputs are  $\hat{x}_{c_k}$ and $\hat{x}_{b_k}$, which are the SSE of received coded symbol ${x_k}$ and information symbol ${b_k}$, and the binary addition operation in the binary encoder is replaced by real-number multiplication in the dual encoder for decoding. 
\end{theorem}


Before presenting the results for the backward decoding \cite{ETT1}, we first define a reverse memory-labeling of a CC code. 


\begin{definition} [Reversed Polynomial]
 Given a generator polynomial $g_f(x)=\frac{a_f(x)}{q_f(x)}=\frac{x^n+a_{n-1}x^{n-1}+\cdots+a_1x+1}{x^n+q_{n-1}x^{n-1}+\cdots+q_1x+1}$, let $g_b(x)=\frac{a_b(x)}{q_b(x)}$ represent the generator polynomial for the \textit{reverse
memory-labeling encoder} of $g_f(x)$, where $a_b(x)={x^n+a_{1}x^{n-1}+\cdots+a_{n-1}x+1}$,  $q_b(x)={x^n+q_{1}x^{n-1}+\cdots+q_{n-1}x+1}$ are obtained respectively by reversing the coefficients of  $a_f(x)$ and $q_f(x)$ and also reversing the register labelling of the encoder, i.e. the $k$-th shift register in the encoder of $g_f(x)$  becomes the $(n-k+1)$-th shift register in the encoder of $g_b(x)$.  $g_b(x)=\frac{a_b(x)}{q_b(x)}$ is called the reversed polynomial of $g_f(x)$.
\end{definition}

\begin{theorem}[Linear Representation of MAP Backward Decoding \cite{ETT1}] For a rate-1 CC with a polynomial of $g_f(x)=\frac{a_f(x)}{q_f(x)}$,
    let $z_f(x)$ be the degree-$l$ MCP of $a_f(x)$. Let 
    \begin{eqnarray} \label{eq10}  \small
    d_f(x) &=& 
    \frac{q_f(x)z_f(x)}{a_f(x)z_f(x)}=\frac{q_f(x)z_f(x)}{x^{n+l}+1} 
    \end{eqnarray}
    be the generator polynomial for the forward decoding. The log-domain SISO backward decoding can be implemented by the dual encoder with the inverse polynomial and reverse memory-labeling of $d_f(x)$, i.e.,
    \begin{eqnarray} \label{eq10a}   \small
     d_b(x) &=& 
    \frac{q_b(x)z_b(x)}{a_b(x)z_b(x)}=\frac{q_b(x)z_b(x)}{x^{n+l}+1} .
    \end{eqnarray} 
\end{theorem}


Fianlly, let ${F_j}[k]$ and ${B_j}[k]$, $j=1, 2, \cdots, n+l$, $k=1, 2,
\cdots, L+n+l$, represent the $j$th shift register content of the dual encoders for forward and backward decoding at time $k$, described by the polynomials  $d_{f}(x)$ and  $d_{b}(x)$. The bidirectional MAP decoding output can be obtained by combining the register contents of the dual encoders for forward and backward decoding as shown in \cite{ETT1}. 

\vspace{-0.7em}
\section{LMAP Decoding: A 4-State RSC Example}  \label{sec::7O5}
\vspace{-0.4em}
Building on the dual encoder structure for rate-1 codes, we now introduce LMAP decoding for rate-1/2 recursive systematic code (RSC)  RSC codes. Consider a rate-1/2 RSC with generator polynomial $\mathbf{g}f(x)=(1, a_f(x)/q_f(x))$, where $a_f(x)=x^n+a_{n-1}x^{n-1}\ldots+a_1x+1$ and $q_f(x)=x^n+q_{n-1}x^{n-1}\ldots+q_1x+1$, with memory length $m=n$. Let $G_a$ and $G_q$ denote $a_f(x)$ and $q_f(x)$ in octal notation. 

For an input binary sequence $\mathbf{b}=(b_1,\ldots,b_L)$ with BPSK modulated symbols $\mathbf{x_b}$, the encoder generates codeword $\mathbf{c}$ where $\mathbf{c}_k=(c_{k}^{1}, c_{k}^{2})$ with $c_{k}^{1}=b_k$. The received sequence is denoted as $\mathbf{y}=(\mathbf{y_1},\ldots,\mathbf{y}_L)$. Let $\hat{x}_{c_{k}^{i}}$ represent the SSE given by $\hat{x}_{c_{k}^{i}} = p_{c_{k}^{i}}(0) - p_{c_{k}^{i}}(1)$ for $i=1,2$, where $p_{c_{k}^{i}}(\ell)$ denotes the a posteriori probability (APP) of $c_{k}^{i}=\ell$.

Let $\hat{x}_{b_k}$ be the SSE of $b_k$ obtained by decoding $\mathbf{y}$, with $\hat{x}_{b_k^1}$ and $\hat{x}_{b_k^2}$ denoting the SSE estimates based on $y_{k}^{1}$ and $y_{k}^{2}$ respectively. Their corresponding LLRs $L_{b_k^1}$ and $L_{b_k^2}$ are given by $
    L_{b_k^1} = \log \frac{1+\hat{x}_{b_k^1}}{1-\hat{x}_{b_k^1}}$ and 
    $L_{b_k^2} = \log \frac{1+\hat{x}_{b_k^2}}{1-\hat{x}_{b_k^2}}$, respectively.

Then, the SSE of the combined LLR $L_{b_k^1} + L_{b_k^2}$, denoted by $\hat{x}_{b_k^1} \circC \hat{x}_{b_k^2}$, is obtained by the following optimal combining:
\begin{equation}  \small
\hat{x}_{b_k^1}\circC\hat{x}_{b_k^2} =  (\hat{x}_{b_k^1} + \hat{x}_{b_k^2})/(1 + \hat{x}_{b_k^1}\hat{x}_{b_k^2})
\end{equation}

\begin{figure} [t]
     \centering
     \begin{subfigure}[b]{0.48\columnwidth}
        \centering
                   \begin{tikzpicture}[auto, node distance=2cm, >=latex']
            \node [inner sep=-1pt] (oplus) at (0, 0) {$\oplus$};
            \node [inner sep=-1pt, above right=0.6cm and 1.45cm of oplus] (oplus1) {$\oplus$};
            \node [inner sep=-1pt, above right=0.6cm and 2.81 cm of oplus] (oplus2) {$\oplus$};
            \node [draw, rectangle, minimum height=0.74cm, minimum width=1cm, right of=oplus, node distance=1cm] (S1) {$M_1$};
            \node [draw, rectangle, minimum height=0.74cm, minimum width=1cm, right of=S1, node distance=1.4cm] (S2) {$M_2$};
            
            \node [left of=oplus, node distance=0.7cm, inner sep=2pt] (input) {$b_k$};
            \node [right of=oplus2, node distance=0.5cm, inner sep=0pt] (output2) {$c_k^2$};
            \node [above of=output2, node distance=0.7cm,inner sep=0pt] (output1) {$c_k^1$};
            
            \draw [->] (input) -- (oplus);
            \draw [->] (oplus) -- (S1) ;
            \draw [->] (S1) -- (S2);
            \draw [->] (S2) -- ++(0.63,0)  |- ($(oplus.west) + (+0.5,-0.75)$) -| (oplus);
            \draw [->] (oplus2) -- (output2);
            \draw [->] (S2.east) -|  (oplus2.south);
            \draw [->] (S1)   -|  (oplus1.south);
            \draw [->] (oplus)-- ++(0.25,0) |-   (oplus1.west);
            \draw [->] (oplus1)-- (oplus2);
            \draw [->] (input) -- ++(0.4,0)  |- (output1);
        \end{tikzpicture}    
            \vspace{-1em}
         \caption{Encoder}

         \label{Fig::7o5::encoder}
     \end{subfigure}
     \hfill
     \begin{subfigure}[b]{0.45\columnwidth}
         \centering
                    \begin{tikzpicture}[
                node distance=0.6cm,
                state/.style={circle, draw, minimum size=0.5cm,inner sep=0pt},
                edge/.style={->, >=stealth', shorten >=1pt, auto, node distance=2cm, semithick}
            ]
            
            \node[state] (S0) {$00$};
            \node[state, below of=S0] (S1) {$01$};
            \node[state, below of=S1] (S2) {$10$};
            \node[state, below of=S2] (S3) {$11$};
            
            \node[state, right of=S0,node distance=3cm] (S0next) {$00$};
            \node[state, below of=S0next] (S1next) {$01$};
            \node[state, below of=S1next] (S2next) {$10$};
            \node[state, below of=S2next] (S3next) {$11$};
            
            \draw[edge] (S0) -- node[above, pos=0.15,sloped, yshift=-1mm, font=\tiny] {$0/00$} (S0next);
            \draw[edge] (S0) -- node[above, pos=0.15,sloped, yshift=-1mm, font=\tiny] {$1/11$} (S2next);
            \draw[edge] (S1) -- node[above, pos=0.15,sloped, yshift=-1mm, font=\tiny] {$0/00$} (S2next);
            \draw[edge] (S1) -- node[above, pos=0.15,sloped, yshift=-1mm, font=\tiny] {$1/11$} (S0next);
            \draw[edge] (S2) -- node[above, pos=0.15,sloped, yshift=-1mm, font=\tiny] {$0/01$} (S1next);
            \draw[edge] (S2) -- node[above, pos=0.15,sloped, yshift=-1mm, font=\tiny] {$1/10$} (S3next);
            \draw[edge] (S3) -- node[above, pos=0.15,sloped, yshift=-1mm, font=\tiny] {$0/01$} (S3next);
            \draw[edge] (S3) -- node[above, pos=0.15,sloped, yshift=-1mm, font=\tiny] {$1/10$} (S1next);
            
            \node[above of=S0, node distance=0.4cm, font=\footnotesize] {Time $k$};
            \node[above of=S0next, node distance=0.4cm, font=\footnotesize] {Time $k+1$};
            
        \end{tikzpicture}
             \vspace{-1em}
        \caption{Trellis}  
        \label{Fig::7o5::trellis}
     \end{subfigure}
     \vspace{-0.5em}
     \caption{The trellis and encoder of $(1, 7_{\mathrm{oct}}/5_{\mathrm{oct}})$}
      \vspace{-1.5em}
     \label{Fig::7o5::trellis&encoder}
\end{figure}
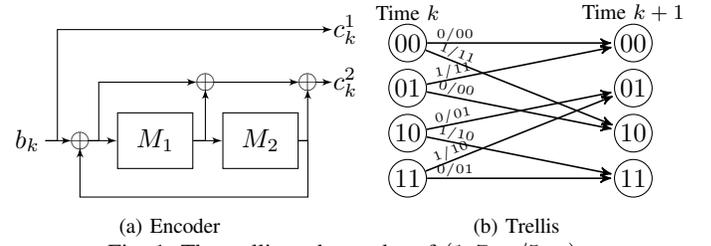

To illustrate the decoding process, we next examine a 4-state RSC with generator polynomial $(1, \frac{1+x+x^2}{1+x^2})$, with octal form $(1,7/5)$. The trellis and encoder structures are shown in Fig. \ref{Fig::7o5::trellis&encoder}, where $M_i[k]$ denotes the $i$-th register value at time $k$. 


We first examine the forward decoding of the BCJR algorithm to reveal recursive relationships that motivate our LMAP decoder. Let $p_{m_k^i}(\ell)$ denote the probability of ${M_i[k] = \ell}$ for $\ell\in\{0,1\}$ given $(\mathbf{y}_1,\ldots,\mathbf{y}_{k-1})$. For forward-only decoding, BCJR determines the soft estimate $\hat{x}_{b_k}^{\rightarrow}$ as
\begin{equation} \label{equ::7o5::BCJR::xb::for}   \small
\begin{split}
       \hat{x}_{b_k}^{\rightarrow} &= (p_{b_k}(0) - p_{b_k}(1))/(p_{b_k}(0) + p_{b_k}(1)), \\
\end{split}
\end{equation}
where $p_{b_k}(0) + p_{b_k}(1)$ serves as normalization: 
\begin{equation}  \label{equ::7o5::BCJR::xb::for::den}   \small
    \begin{split}
       p_{b_k}(0) \!+\! p_{b_k}(1) & = (\alpha_{k}^0 + \alpha_{k}^1)(p_{\mathbf{c}_{k}}(00)+ p_{\mathbf{c}_{k}}(11)) \\
       & + (\alpha_{k}^2 + \alpha_{k}^3)(p_{\mathbf{c}_{k}}(01) + p_{\mathbf{c}_{k}}(10)) \\
       &  = \frac{1}{2} + \frac{1}{2}\hat{x}_{c_k^1} \hat{x}_{c_k^2} \left(\alpha_{k}^0 \!+\! \alpha_{k}^1 \!-\! \alpha_{k}^2 \!-\! \alpha_{k}^3  \right),
    \end{split}
\end{equation}
and similarly,
\begin{equation} \label{equ::7o5::BCJR::xb::for::num}   \small
    \begin{split}
       p_{b_k}(0) \!-\! p_{b_k}(1) =\frac{1}{2}\hat{x}_{c_k^1} + \frac{1}{2} \hat{x}_{c_k^2} \left(\alpha_{k}^0 \!+\! \alpha_{k}^1 \!-\! \alpha_{k}^2 \!-\! \alpha_{k}^3  \right).
    \end{split}
\end{equation}

Equations \eqref{equ::7o5::BCJR::xb::for::num} and \eqref{equ::7o5::BCJR::xb::for::den} are derived using relationships $\frac{1}{2}(\hat{x}_{c_k^1}\hat{x}_{c_k^2} +1) = p_{\mathbf{c}_{k}}(00)+p_{\mathbf{c}_{k}}(11)$ and $\frac{1}{2}(1 - \hat{x}_{c_k^1} \hat{x}_{c_k^2}) =  p_{\mathbf{c}_{k}}(01) +  p_{\mathbf{c}_{k}}(10)$.  Furthermore, we observe from the trellis that $p_{m_k^1}(0)=\alpha_{k}^0 + \alpha_{k}^1$ and $p_{m_k^1}(1)=\alpha_{k}^2 + \alpha_{k}^3$. Thus, let  $\hat{x}_{m_k^1}$=$p_{m_k^1}(0)-p_{m_k^1}(1)$ denote the SSE of $M_1[k]$, and we have
\begin{equation}  \label{equ::7o5::BCJR::xb::for2}   \small
     \hat{x}_{b_k}^{\rightarrow} = \frac{1}{\lambda_k}\left(\hat{x}_{c_k^1} +\hat{x}_{c_k^2}\hat{x}_{m_k^1}\right) = \hat{x}_{c_k^1} \circC(\hat{x}_{c_k^2}\hat{x}_{m_k^1}),
\end{equation}
where $\lambda_k = 1 + \hat{x}_{c_k^1}\hat{x}_{c_k^2} \hat{x}_{m_k^1}$ is the normalization factor.

\begin{remark}
Equation \eqref{equ::7o5::BCJR::xb::for2} resembles the encoder structure in reverse. The decoder's SSE combining $\hat{x}_{b_k} =\hat{x}_{c_k^1} \circC(\hat{x}_{c_k^2}\hat{x}_{m_k^1})$ mirrors the encoder's operations $b_k = c_k^1$ and $b_k = c_k^2 \oplus M_1[k]$. This reveals a duality between encoding and decoding.
\end{remark}

Observing the trellis diagram in Fig. \ref{Fig::7o5::trellis}, we see $\hat{x}_{m_{k+1}^1}=(\alpha_{k+1}^0 + \alpha_{k+1}^1) - (\alpha_{k+1}^2 + \alpha_{k+1}^3)$.  Following the recursive updates in BCJR, $\hat{x}_{m_{k+1}^1}$ can be obtained as:
\begin{equation}  \label{equ::7o5::m1update2}  \small
    \hat{x}_{m_{k+1}^1} = \frac{1}{\lambda_k}\left(\hat{x}_{c_k^1}\hat{x}_{m_k^{2}}+\hat{x}_{c_k^2}\hat{x}_{m_k^{12}}\right),
\end{equation}
where $\hat{x}_{m_k^2} = \alpha_{k}^0 - \alpha_{k}^1 + \alpha_{k}^2 - \alpha_{k}^3$ is the SSE of $M_2[k]$, and $\hat{x}_{m_k^{12}} = \alpha_{k}^0 - \alpha_{k}^1 + \alpha_{k}^2 - \alpha_{k}^3$ represents the joint SSE of $M_1[k]\oplus M_2[k]$. Similarly, we can obtain:
\begin{equation}  \label{equ::7o5::m2update}   \small
    \hat{x}_{m_{k+1}^2} = \frac{1}{\lambda_k}\left(\hat{x}_{c_k^1}\hat{x}_{c_k^2}+\hat{x}_{m_k^{1}} \right)
\end{equation}
\begin{equation}  \label{equ::7o5::m12update}   \small
    \hat{x}_{m_{k+1}^{12}} = \frac{1}{\lambda_k}\left(\hat{x}_{c_k^1}\hat{x}_{m_{k}^{12}}+ \hat{x}_{c_k^2}\hat{x}_{m_k^{2}} \right).
\end{equation}

All memory SSEs ($\hat{x}_{m_{1}^1}$, $\hat{x}_{m_{1}^2}$, $\hat{x}_{m_{1}^{12}}$) are initialized to 1 at $k=1$, following from the initial BCJR conditions where $\alpha_1^0 = 1$ and $\alpha_1^i = 0$ for $i \in\{1,2,3\}$. 



%


\begin{figure} [t]
     \centering
     \hspace{-0.81em}
     \begin{subfigure}[b]{\columnwidth}
         \centering
                   \begin{tikzpicture}[auto, node distance=2cm, >=latex']
           
            \node [draw, rectangle, minimum height=0.7cm, minimum width=1cm] (F2_1) at (0, 0) {$F_2^{(1)}$};
            \node [draw, rectangle, minimum height=0.7cm, minimum width=1cm, right of=F2_1, node distance=2cm] (F1_1) {$F_{1}^{(1)}$};
            \node [draw, rectangle, minimum height=0.7cm, minimum width=1cm, right of=F1_1, node distance=3.7cm] (F12_1) {$F_{12}^{(1)}$};

            \node [inner sep=-1pt, left of=F2_1, node distance=1cm] (otF2_1)  {$\otimes$};
            \node [inner sep=-1pt, left of=F1_1, node distance=1cm] (otF1_1) {$\otimes$};
             \node [inner sep=-1pt, left of=F12_1, node distance=1cm] (otF12_1) {$\otimes$};
            \node [inner sep=-1pt, right of=F1_1, node distance=1cm] (otOut1) {$\otimes$};

            \node [above left=-0.1cm and 0.65 cm of F2_1] (input1_1) {$\hat{x}_{c_k^1}$};
            \node [below left=-0.1cm and 0.65 cm of F2_1] (input1_2) {$\hat{x}_{c_k^2}$};
            \node [right of=otOut1, node distance=0.8cm] (out1) {$\hat{x}_{b_k}^{\rightarrow}$};

             \node [inner sep=-0.5pt, above =0.3cm  of otF12_1] (input1_3) {$\hat{x}_{c_k^1}$};
            
            \draw [->] (otF2_1) -- (F2_1);
            \draw [->] (input1_1) -| (otF2_1);
            \draw [->] (input1_2) -| (otF2_1);
            \draw [-] (F2_1) -- (otF1_1);
            \draw [->] (otF1_1) -- (F1_1);
            \draw [->] (input1_1) -| (otF1_1);
            \draw [->] (F12_1.east) -- ++(0.2,0)  |- ($(F12_1.west) + (0,-0.6)$) -| (otF12_1.south); 
            \draw [->] (otF12_1) -- (F12_1);
            \draw [->] (input1_3) -- (otF12_1);
            \draw [-] (F1_1) -- (otOut1);
            \draw [->] (otOut1) -- (out1);
            \draw [->] (input1_2) -| (otOut1);

        \end{tikzpicture}    
            \vspace{-1em}
        \caption{Forward decoding $\mathrm{DF}_1$}  
        \vspace{-0em}
        \label{Fig::general::decoder::forward}
     \end{subfigure}
     \hfill
      \begin{subfigure}[b]{\columnwidth}
         \centering
                   \begin{tikzpicture}[auto, node distance=2cm, >=latex']

            \node [draw, rectangle, minimum height=0.7cm, minimum width=1cm] (F2_2) at (0, 0) {$F_2^{(2)}$};
            \node [draw, rectangle, minimum height=0.7cm, minimum width=1cm, right of=F2_2, node distance=2cm] (F12_2) {$F_{12}^{(2)}$};
            \node [draw, rectangle, minimum height=0.7cm, minimum width=1cm, right of=F12_2, node distance=2cm] (F1_2) {$F_{1}^{(2)}$};
            
            \node [inner sep=-1pt, right of=F2_2, node distance=1cm] (otF2_2)  {$\otimes$};
            \node [inner sep=-1pt, right of=F12_2, node distance=1cm] (otF12_2) {$\otimes$};
            \node [inner sep=-1pt, right of=F1_2, node distance=1.1cm] (otOut2) {$\otimes$};

            \node [above left=-0.1cm and 0.6 cm of F2_2] (input2_2) {$\hat{x}_{c_k^2}$};
            \node [right of=otOut2, node distance=0.8cm] (out2) {$\hat{x}_{b_k}^{\rightarrow}$};
            
            \draw [->] (F1_2.east) -- ++(0.2,0)  |- ($(F2_2.west) + (-0.3,-0.6)$) |- (F2_2.west); 
            \draw [-] (F2_2) -- (otF2_2);
            \draw [->] (otF2_2) -- (F12_2);
            \draw [-] (F12_2) -- (otF12_2);
            \draw [->] (otF12_2) -- (F1_2);
            \draw [->] (input2_2) -| (otF2_2);
            \draw [->] (input2_2) -| (otF12_2);
            \draw [-] (F1_2) -- (otOut2);
            \draw [->] (otOut2) -- (out2);
            \draw [->] (input2_2) -| (otOut2);

        \end{tikzpicture}    
        \caption{Forward decoding $\mathrm{DF}_2$}
        \vspace{-0em}
        \label{Fig::general::decoder::forward}
     \end{subfigure}
    \vspace{-1.5em}
     \caption{The LMAP decoder (forward decoding) of $(1, 7_{\mathrm{oct}}/5_{\mathrm{oct}})$}
    \vspace{-1.5em}
     \label{Fig::7o5::decoder}
\end{figure}
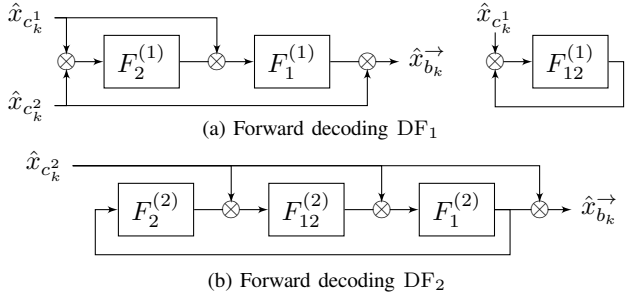

For backward decoding, let $\hat{x}_{s_{k}^i}$ denote the SSE of memory state $M_i[k]$ based on future received symbols $(\mathbf{y}_k,\mathbf{y}_{k+1},\ldots,\mathbf{y}_L)$, and $\hat{x}_{s_{k}^{12}}$ represent the joint SSE of $M_1[k]\oplus M_2[k]$. Similar recursive relationships exist:
\begin{equation}   \label{equ::7o5::s1update}  \small
    \hat{x}_{s_{k}^1} = \frac{1}{\rho_{k}}\left( x_{c_{k}^1}x_{c_{k}^2} +  \hat{x}_{s_{k+1}^2} \right),
\end{equation}
\begin{equation}   \label{equ::7o5::s2update} \small
    \hat{x}_{s_{k}^2} = \frac{1}{\rho_{k}}\left( x_{c_{k}^1}\hat{x}_{s_{k+1}^1} +  x_{c_{k}^2}\hat{x}_{s_{k+1}^{12}} \right),
\end{equation}
\begin{equation}    \label{equ::7o5::s12update} \small
    \hat{x}_{s_{k}^{12}} = \frac{1}{\rho_{k}}\left( x_{c_{k}^1} \hat{x}_{s_{k+1}^{12}}+  x_{c_{k}^2}\hat{x}_{s_{k+1}^{1}} \right),
\end{equation}
where $\rho_k = 1 + x_{c_{k}^1}x_{c_{k}^2}\hat{x}_{s_{k+1}^2}$ is the normalization factor. These recursive relationships are obtained using of recursive updates of $\beta$ values in the BCJR algorithm.
\begin{remark}
    The forward recursive updates can be implemented using two forward decoders $\mathrm{DF}1$ and $\mathrm{DF}2$ as shown in Fig. \ref{Fig::7o5::decoder}. Each register $F_1^{(1)}$, $F_2^{(1)}$, and $F_{12}^{(1)}$ store SSEs $\hat{x}_{m_k^1}$, $\hat{x}_{m_k^2}$, and $\hat{x}_{m_k^{12}}$ respectively. $\mathrm{DF}_1$ calculates the first terms in equations \eqref{equ::7o5::m1update2}--\eqref{equ::7o5::m12update} using both $\hat{x}_{c_k^1}$ and $\hat{x}_{c_k^2}$, while $\mathrm{DF}_2$ computes the second terms using only $\hat{x}_{c_k^2}$. The register values from both decoders are then optimally combined to obtain the complete memory SSE updates. The backward recursive updates can be implemented similarly.
\end{remark}

While \eqref{equ::7o5::BCJR::xb::for2} obtains single-direction outputs, achieving MAP performance requires combining probabilities from both directions. For the $(1,\frac{7}{5})$ RSC, the final bidirectional LLR is:
\begin{equation} \label{equ::biLLR}   \small
L_{b_k} = \log\frac{(1+\hat{x}_{c_k^1})(\delta_k + \mu_k)}{(1-\hat{x}_{c_k^1})(\delta_k - \mu_k)},
\end{equation}
where
\begin{equation} \notag  \small
   \delta_k = 1 + \hat{x}_{m_k^1}\hat{x}_{s_{k+1}^2} + \hat{x}_{c_k^2}\hat{x}_{m_k^2}\hat{x}_{s_{k+1}^{12}} + \hat{x}_{c_k^2}\hat{x}_{m_k^{12}}\hat{x}_{s_{k+1}^{1}},
\end{equation}
\begin{equation}  \notag  \small
   \mu_k = \hat{x}_{c_k^2}\hat{x}_{m_k^1} + \hat{x}_{c_k^2}\hat{x}_{s_{k+1}^{2}} + \hat{x}_{m_k^{12}}\hat{x}_{s_{k+1}^{12}} + \hat{x}_{m_k^2}\hat{x}_{s_{k+1}^1}.
\end{equation}
All memory SSEs ($\hat{x}_{m_k^i}$ and $\hat{x}_{s_{k+1}^i}$) are directly available by recursive updating \eqref{equ::7o5::m1update2}--\eqref{equ::7o5::s12update}. This output is mathematically equivalent to the bidirectional BCJR algorithm. 

\vspace{-0.8em}
\section{LMAP Decoding of A General Rate 1/2 RSC}  \label{sec::GeneralRSC}
\vspace{-0.3em}
\subsection{LMAP Decoder Structure}
\begin{figure} [t]
     \centering
     \hspace{-0.81em}
     \begin{subfigure}[b]{\columnwidth}
         \centering
                   \begin{tikzpicture}[auto, node distance=2cm, >=latex']
           
            \node [draw, rectangle, minimum height=0.7cm, minimum width=0.8cm, inner sep=0pt] (F1_1) at (0, 0) {$F_{\mathcal{J}_1}^{(1)}$};
            \node [draw, rectangle, minimum height=0.7cm, minimum width=0.8cm, inner sep=0pt, right of=F1_1, node distance=1.35cm] (F2_1) {$F_{\mathcal{J}_2}^{(1)}$};
            \node [inner sep=0pt, right of=F2_1, node distance=1.2cm] (dotsB1) {$\cdots$};
            \node [draw, rectangle, minimum height=0.7cm, minimum width=0.8cm, inner sep=0pt, right of=dotsB1, node distance=1.2cm] (FN-1_1) {$F_{\!\mathcal{J}_{\!N\!-\!2}}^{(1)}$};

            \node [draw, rectangle, minimum height=0.7cm, minimum width=0.8cm, inner sep=0pt, right of=FN-1_1, node distance=3.2cm] (FS) {$F_{\mathcal{S}}^{(1)}$};

            \node [inner sep=-1pt, left of=F1_1, node distance=0.75cm] (otF0_1)  {$\otimes$};
            \node [inner sep=-1pt, right of=F1_1, node distance=0.6cm] (otF1_1) {$\otimes$};
             \node [inner sep=-1pt, right of=F2_1, node distance=0.6cm] (otF2_1) {$\otimes$};
            \node [inner sep=-1pt, left of=FN-1_1, node distance=0.75cm] (otFN-1_1) {$\otimes$};
             \node [inner sep=-1pt, right of=FN-1_1, node distance=0.7cm] (otFN_1) {$\otimes$};

             \node [inner sep=-1pt, left of=FS, node distance=0.75cm] (otF0_S)  {$\otimes$};

            \node [ inner sep=1pt, below =0.4cm of otF1_1] (f1_1) {\small {$d_1^{(1)}$}};
            \node [ inner sep=1pt, below =0.4cm of otF2_1] (f2_1) {\small {$d_2^{(1)}$}};
            \node [ inner sep=1pt, below =0.4cm of otFN-1_1] (fN-1_1) {\small {$d_{N-3}^{(1)}$}};
            \node [inner sep=0pt, below of=dotsB1, node distance=1.183cm] (dotsd1) {$\cdots$};
            \node [inner sep=0pt, above of=dotsB1, node distance=0.583cm] (dotso1) {$\cdots$};

            \node [ inner sep=1pt, below =0.5cm of otF0_S] (fs) {\small {$d_{s}$}};
            
            \node [above left=0cm and 0.45 cm of F1_1,inner sep=0.5pt] (input1_1) {$\hat{x}_{c_k^1}$};
            \node [below left=0.6cm and 0.45 cm of F1_1,inner sep=0.5pt] (input1_2) {$\hat{x}_{c_k^2}$};
            \node [right of=otFN_1, node distance=0.6cm,inner sep=0.5pt] (out1) {$\hat{x}_{b_k}^{\rightarrow}$};
            \node [above left=0.0cm and 0.45 cm of FS,inner sep=0.5pt] (inputS_1) {$\hat{x}_{c_k^1}$};
            \node [below left=0.6cm and 0.45 cm of FS,inner sep=0.5pt] (inputS_2) {$\hat{x}_{c_k^2}$};
            
            \draw [->] (input1_1) -| (otF0_1);
            \draw [->] (input1_2) -| (otF0_1);

            \draw [-] (input1_2) -| (f1_1);
            \draw [-] (input1_2) -| (f2_1);
            \draw [->] (f2_1) |- (dotsd1);
            \draw [-] (dotsd1) -| (fN-1_1);
            \draw [->] (dotsd1) -| (otFN_1);
            \draw [->] (f1_1) -- (otF1_1);
            \draw [->] (f2_1) -- (otF2_1);
            \draw [->] (fN-1_1) -- (otFN-1_1);

            \draw [->] (input1_1) -| (otF1_1);
            \draw [->] (input1_1) -| (otF2_1);
            \draw [->] (otF2_1) |- (dotso1);
            \draw [->] (dotso1) -| (otFN-1_1);

            \draw [->] (otF0_1) -- (F1_1);
            \draw [->] (otF1_1) -- (F2_1);
            \draw [->] (otF2_1) -- (dotsB1);
            \draw [->] (otFN-1_1) -- (FN-1_1);
            \draw [->] (FN-1_1) -- (otFN_1);

            \draw [-] (F1_1) -- (otF1_1);
            \draw [-] (F2_1) -- (otF2_1);
            \draw [-] (dotsB1) -- (otFN-1_1);
            \draw [->] (otFN_1) -- (out1);

            \draw [-] (inputS_2) -| (fs);
            \draw [->] (fs) -- (otF0_S);
            \draw [->] (inputS_1) |- (otF0_S);
             \draw [->] (FS.east) -| ++(0.2,0.56)  -| (otF0_S); 
             \draw [->] (otF0_S) -- (FS);

        \end{tikzpicture}    
            \vspace{-1.5em}
        \caption{Forward decoding $\mathrm{DF}_1$}  
        \vspace{-0em}
        \label{Fig::general::decoder::forward}
     \end{subfigure}
     \hfill
      \begin{subfigure}[b]{\columnwidth}
         \centering
                   \begin{tikzpicture}[auto, node distance=2cm, >=latex']
           
            \node [draw, rectangle, minimum height=0.7cm, minimum width=0.8cm, inner sep=0pt] (F1_1) at (0, 0) {$F_{\mathcal{I}_1}^{(2)}$};
            \node [draw, rectangle, minimum height=0.7cm, minimum width=0.8cm, inner sep=0pt, right of=F1_1, node distance=1.4cm] (F2_1) {$F_{\mathcal{I}_2}^{(2)}$};
            \node [inner sep=0pt, right of=F2_1, node distance=1.2cm] (dotsB1) {$\cdots$};
            \node [draw, rectangle, minimum height=0.7cm, minimum width=0.8cm, inner sep=0pt, right of=dotsB1, node distance=1.3cm] (FN-1_1) {$F_{\!\mathcal{I}_{\!N\!-\!2}}^{(2)}$};
            \node [draw, rectangle, minimum height=0.7cm, minimum width=0.8cm, inner sep=0pt, right of=FN-1_1, node distance=1.4cm] (FN_1) {$F_{\!\mathcal{I}_{\!N\!-\!1}}^{(2)}$};

            \node [inner sep=-1pt, right of=F1_1, node distance=0.65cm] (otF1_1) {$\otimes$};
             \node [inner sep=-1pt, right of=F2_1, node distance=0.6cm] (otF2_1) {$\otimes$};
             \node [inner sep=-1pt, left of=FN-1_1, node distance=0.8cm] (otFN-2_1) {$\otimes$};
            \node [inner sep=-1pt, left of=FN_1, node distance=0.75cm] (otFN-1_1) {$\otimes$};
             \node [inner sep=-1pt, right of=FN_1, node distance=0.8cm] (otFN_1) {$\otimes$};


            \node [ inner sep=1pt, above =0.4cm of otF1_1] (f1_1) {\small {$d_{N-2}^{(2)}$}};
            \node [ inner sep=1pt, above =0.4cm of otF2_1] (f2_1) {\small {$d_{N-3}^{(2)}$}};
            \node [ inner sep=1pt, above =0.4cm of otFN-2_1] (fN-2_1) {\small {$d_{2}^{(2)}$}};
            \node [ inner sep=1pt, above =0.4cm of otFN-1_1] (fN-1_1) {\small {$d_{1}^{(2)}$}};
            \node [inner sep=0pt, above of=dotsB1, node distance=1.293cm] (dotsd1) {$\cdots$};

            \node [above left=0.6cm and 0.4 cm of F1_1] (input1_2) {$\hat{x}_{c_k^2}$};
            \node [right of=otFN_1, node distance=0.65cm, inner sep=0pt] (out1) {$\hat{x}_{b_k}^{\rightarrow}$};
            

            \draw [-] (input1_2) -| (f1_1);
            \draw [-] (input1_2) -| (f2_1);
            \draw [->] (f2_1) |- (dotsd1);
            \draw [-] (dotsd1) -| (fN-1_1);
            \draw [-] (dotsd1) -| (fN-2_1);
            \draw [->] (dotsd1) -| (otFN_1);
            \draw [->] (f1_1) -- (otF1_1);
            \draw [->] (f2_1) -- (otF2_1);
            \draw [->] (fN-1_1) -- (otFN-1_1);
            \draw [->] (fN-2_1) -- (otFN-2_1);

            \draw [->] (otF1_1) -- (F2_1);
            \draw [->] (otF2_1) -- (dotsB1);
            \draw [->] (otFN-1_1) -- (FN_1);
            \draw [-] (F1_1) -- (otF1_1);
            \draw [-] (F2_1) -- (otF2_1);
            \draw [-] (dotsB1) -- (otFN-2_1);
            \draw [->] (otFN-2_1) -- (FN-1_1);
            \draw [-] (FN-1_1) -- (otFN-1_1);
            \draw [-] (FN_1) -- (otFN_1);
            \draw [->] (otFN_1) -- (out1);

            \draw [->] (FN_1.east) -| ++(0.1,-0.56)   -| ($(F1_1.west) + (-0.4,0)$) |- (F1_1);

        \end{tikzpicture}    
        \caption{Forward decoding $\mathrm{DF}_2$}
        \vspace{-0em}
        \label{Fig::general::decoder::forward}
     \end{subfigure}
     \hfill
     \begin{subfigure}[b]{\columnwidth}
        \centering
                   \begin{tikzpicture}[auto, node distance=2cm, >=latex']
           
            \node [draw, rectangle, minimum height=0.74cm, minimum width=0.8cm, inner sep=0pt] (F1_1) at (0, 0) {$B_{\mathcal{J}_1}^{(1)}$};
            \node [inner sep=0pt, right of=F1_1, node distance=1.2cm] (dotsB1) {$\cdots$};
            \node [draw, rectangle, minimum height=0.74cm, minimum width=0.8cm, inner sep=0pt, right of=dotsB1, node distance=1.2cm] (F2_1) {$B_{\!\mathcal{J}_{\!N\!-\!3}}^{(1)}$};
            \node [draw, rectangle, minimum height=0.74cm, minimum width=0.8cm, inner sep=0pt, right of=F2_1, node distance=1.4cm] (FN-1_1) {$B_{\!\mathcal{J}_{\!N\!-\!2}}^{(1)}$};

            \node [draw, rectangle, minimum height=0.74cm, minimum width=0.8cm, inner sep=0pt, right of=FN-1_1, node distance=2.2cm] (FS) {$B_{\mathcal{S}}^{(1)}$};

            \node [inner sep=-1pt, left of=F1_1, node distance=0.65cm] (otF0_1)  {$\otimes$};
            \node [inner sep=-1pt, right of=F1_1, node distance=0.7cm] (otF1_1) {$\otimes$};
             \node [inner sep=-1pt, left of=F2_1, node distance=0.65cm] (otF2_1) {$\otimes$};
            \node [inner sep=-1pt, right of=F2_1, node distance=0.75cm] (otFN-1_1) {$\otimes$};
             \node [inner sep=-1pt, right of=FN-1_1, node distance=0.75cm] (otFN_1) {$\otimes$};

             \node [inner sep=-1pt, right of=FS, node distance=0.7cm] (otF0_S)  {$\otimes$};

            \node [ inner sep=1pt, below =0.4cm of otF1_1] (f1_1) {\small {$d_1^{(1)}$}};
            \node [ inner sep=1pt, below =0.4cm of otF2_1] (f2_1) {\small {$d_{N-4}^{(1)}$}};
            \node [ inner sep=1pt, below =0.4cm of otFN-1_1] (fN-1_1) {\small {$d_{N-3}^{(1)}$}};
            \node [inner sep=0pt, below of=dotsB1, node distance=1.238cm] (dotsd1) {$\cdots$};
            \node [inner sep=0pt, above of=dotsB1, node distance=0.635cm] (dotso1) {$\cdots$};

            \node [ inner sep=1pt, below =0.4cm of otF0_S] (fs) {\small {$d_{s}$}};
            
            \node [above right=0cm and 0.5 cm of FN-1_1, inner sep=1pt] (input1_1) {$\hat{x}_{c_k^1}$};
            \node [below right=0.6cm and 0.5 cm of FN-1_1, inner sep=1pt] (input1_2) {$\hat{x}_{c_k^2}$};
            \node [left of=otF0_1, node distance=0.6cm, inner sep=1pt] (out1) {$\hat{x}_{b_k}^{\rightarrow}$};
            \node [above right=0cm and 0.35 cm of FS, inner sep=1pt] (inputS_1) {$\hat{x}_{c_k^1}$};
            \node [below right=0.6cm and 0.35 cm of FS, inner sep=1pt] (inputS_2) {$\hat{x}_{c_k^2}$};
            
            \draw [->] (input1_1) -| (otFN_1);
            \draw [->] (input1_2) -| (otFN_1);

            \draw [-] (input1_2) -| (fN-1_1);
            \draw [-] (input1_2) -| (f2_1);
            \draw [->] (f2_1) |- (dotsd1);
            \draw [-] (dotsd1) -| (f1_1);
            \draw [->] (dotsd1) -| (otF0_1);
            \draw [->] (f1_1) -- (otF1_1);
            \draw [->] (f2_1) -- (otF2_1);
            \draw [->] (fN-1_1) -- (otFN-1_1);

            \draw [->] (input1_1) -| (otFN-1_1);
            \draw [->] (input1_1) -| (otF2_1);
            \draw [->] (otF2_1) |- (dotso1);
            \draw [->] (dotso1) -| (otF1_1);

            \draw [->] (otFN_1) -- (FN-1_1);
            \draw [->] (otFN-1_1) -- (F2_1);
            \draw [->] (otF2_1) -- (dotsB1);
            \draw [-] (F2_1) -- (otF2_1);
            \draw [-] (FN-1_1) -- (otFN-1_1);
            \draw [-] (dotsB1) -- (otF1_1);
            \draw [->] (otF1_1) -- (F1_1);            
            \draw [-] (F1_1) -- (otF0_1);
            \draw [->] (otF0_1) -- (out1);

            \draw [-] (inputS_2) -| (fs);
            \draw [->] (fs) -- (otF0_S);
            \draw [->] (inputS_1) |- (otF0_S);
             \draw [->] (FS.west) -| ++(-0.2,0.56)  -| (otF0_S); 
             \draw [->] (otF0_S) -- (FS);

        \end{tikzpicture}    
            \vspace{-1.5em}
         \caption{Backward decoding $\mathrm{DB}_1$}
          \vspace{-0em}
         \label{Fig::general::decoder::backward}
     \end{subfigure}
     \hfill
     \begin{subfigure}[b]{\columnwidth}
        \centering
                   \begin{tikzpicture}[auto, node distance=2cm, >=latex']
           
            \node [draw, rectangle, minimum height=0.7cm, minimum width=0.8cm, inner sep=0pt] (F1_1) at (0, 0) {$B_{\mathcal{I}_1}^{(2)}$};
            \node [draw, rectangle, minimum height=0.7cm, minimum width=0.8cm, inner sep=0pt, right of=F1_1, node distance=1.45cm] (F2_1) {$B_{\mathcal{I}_2}^{(2)}$};
            \node [inner sep=0pt, right of=F2_1, node distance=1.2cm] (dotsB1) {$\cdots$};
            \node [draw, rectangle, minimum height=0.7cm, minimum width=0.8cm, inner sep=0pt, right of=dotsB1, node distance=1.3cm] (FN-1_1) {$B_{\mathcal{I}_{N-2}}^{(2)}$};
            \node [draw, rectangle, minimum height=0.7cm, minimum width=0.8cm, inner sep=0pt, right of=FN-1_1, node distance=1.6cm] (FN_1) {$B_{\mathcal{I}_{N-1}}^{(2)}$};

            \node [inner sep=-1pt, right of=F1_1, node distance=0.8cm] (otF1_1) {$\otimes$};
             \node [inner sep=-1pt, right of=F2_1, node distance=0.7cm] (otF2_1) {$\otimes$};
             \node [inner sep=-1pt, left of=FN-1_1, node distance=0.75cm] (otFN-2_1) {$\otimes$};
            \node [inner sep=-1pt, left of=FN_1, node distance=0.75cm] (otFN-1_1) {$\otimes$};
             \node [inner sep=-1pt, left of=F1_1, node distance=0.75cm] (otFN_1) {$\otimes$};


            \node [ inner sep=1pt, above =0.4cm of otF1_1] (f1_1) {\small {$d_{N-2}^{(2)}$}};
            \node [ inner sep=1pt, above =0.4cm of otF2_1] (f2_1) {\small {$d_{N-3}^{(2)}$}};
            \node [ inner sep=1pt, above =0.4cm of otFN-2_1] (fN-2_1) {\small {$d_{2}^{(2)}$}};
            \node [ inner sep=1pt, above =0.4cm of otFN-1_1] (fN-1_1) {\small {$d_{1}^{(2)}$}};
            \node [inner sep=0pt, above of=dotsB1, node distance=1.28cm] (dotsd1) {$\cdots$};

            \node [above right=0.6cm and 0.3 cm of FN_1] (input1_2) {$\hat{x}_{c_k^2}$};
            \node [left of=otFN_1, node distance=0.8cm] (out1) {$\hat{x}_{b_k}^{\rightarrow}$};
            

            \draw [-] (input1_2) -| (fN-1_1);
            \draw [-] (input1_2) -| (fN-2_1);
            \draw [->] (fN-2_1) |- (dotsd1);
            \draw [-] (dotsd1) -| (f1_1);
            \draw [-] (dotsd1) -| (f2_1);
            \draw [->] (dotsd1) -| (otFN_1);
            \draw [->] (f1_1) -- (otF1_1);
            \draw [->] (f2_1) -- (otF2_1);
            \draw [->] (fN-1_1) -- (otFN-1_1);
            \draw [->] (fN-2_1) -- (otFN-2_1);

            \draw [-] (otF1_1) -- (F2_1);
            \draw [-] (otF2_1) -- (dotsB1);
            \draw [-] (otFN-1_1) -- (FN_1);
            \draw [<-] (F1_1) -- (otF1_1);
            \draw [<-] (F2_1) -- (otF2_1);
            \draw [<-] (dotsB1) -- (otFN-2_1);
            \draw [-] (otFN-2_1) -- (FN-1_1);
            \draw [<-] (FN-1_1) -- (otFN-1_1);
            \draw [-] (F1_1) -- (otFN_1);
            \draw [->] (otFN_1) -- (out1);

            \draw [->] (F1_1.west) -| ++(-0.1,-0.56)   -| ($(FN_1.east) + (0.4,0)$) |- (FN_1);

        \end{tikzpicture}    
            \vspace{-1em}
         \caption{Backward decoding $\mathrm{DB}_2$}
        \vspace{0em}
         \label{Fig::general::decoder::backward}
     \end{subfigure}
    \vspace{-1em}
     \caption{The LMAP decoder of a general rate-$\frac{1}{2}$ RSC}
    \vspace{-1.5em}
     \label{Fig::general::decoder}
\end{figure}
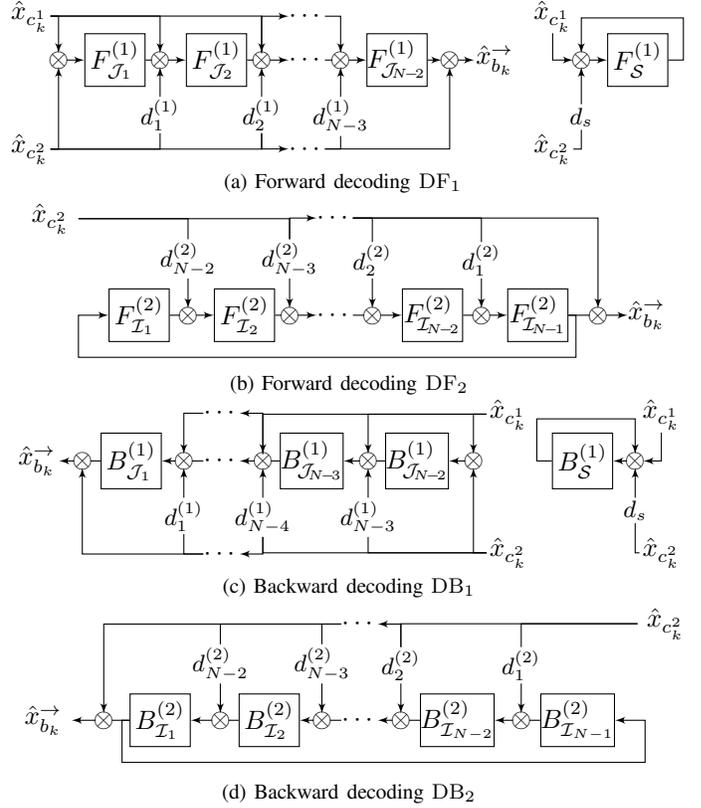

\vspace{-0.3em}
Consider a rate-1/2 RSC with generator polynomial $\mathbf{g}(x) = (1, a_f(x)/q_f(x))$, where $a_f(x)$ is a primitive polynomial of degree $n$. The LMAP decoder (Fig. \ref{Fig::general::decoder}) consists of four shift register modules: forward decoders ($\mathrm{DF}_1$, $\mathrm{DF}_2$) and backward decoders ($\mathrm{DB}_1$, $\mathrm{DB}_2$), each containing $N-1$ registers, where $N$ is the number of states. The decoder is fully defined by:
$$\small \mathcal{D} = \{d_{s}, d_{f_1}(x), d_{f_2}(x), \mathcal{S}, \mathcal{I}, \mathcal{J}\}$$
where $d_s$ is the coefficient for self-updating register (SUR) connection, and 
$$\small d_{f_1}(x) = 1+d_1^{(1)}x+d_2^{(1)}x^2+\ldots+ d_{N-3}^{(1)}x^{N-3} + x^{N-2}$$
$$\small d_{f_2}(x) = 1+d_1^{(2)}x+d_2^{(2)}x^2+\ldots+ d_{N-2}^{(2)}x^{N-2} + x^{N-1}$$
are the decoder polynomials. These polynomials define the connection coefficients for $\hat{x}_{c_k^2}$ in $\mathrm{DF}_1$ and $\mathrm{DF}2$ respectively.

Each register in $\mathrm{DF}_2$ and $\mathrm{DB}_2$ is labeled with $\mathcal{I}_j \in \mathcal{I} = (\mathcal{I}_1, \mathcal{I}_2, \dots, \mathcal{I}_{N-1})$, while each register in $\mathrm{DF}_1$ and $\mathrm{DB}_1$ is labeled with $\mathcal{J}_i \in \mathcal{J} =  (\mathcal{J}_1, \mathcal{J}_2, \dots, \mathcal{J}_{N-2})$. Particularly, $\mathcal{S}$ is the label of SUR.

\begin{remark}
The LMAP decoder structure has three key characteristics: (i) the SURs in $\mathrm{DF}_1$ and $\mathrm{DB}_1$ update recursively from themselves, (ii) $\mathrm{DF}_1$ and $\mathrm{DF}_2$ share the same start and end register labels (excluding SUR), with $\mathcal{J}_1 = \mathcal{I}_1$ and $\mathcal{J}_{N-2} = \mathcal{I}_{N-1}$ (similarly for $\mathrm{DB}_1$ and $\mathrm{DB}_2$), and (iii) $\hat{x}_{c_k^1}$ connects to all registers in $\mathrm{DF}_1$ and $\mathrm{DB}_1$, while $\hat{x}_{c_k^2}$ connects only to selected registers with coefficients defined by $d_{f_1}(x)$ and $d_{f_2}(x)$ respectively.
\end{remark}

The decoder parameters $\mathcal{D}$ are determined offline. Since $a_f(x)$ is primitive of degree $n$, it divides $x^{N-1}+1$, allowing us to define the complementary polynomial:
\begin{equation} \small
z_f(x) = (x^{N-1}+1)/a_f(x)
\end{equation}
The decoder polynomials are then obtained as:
\begin{equation} \small
d_{f_2}(x) = z_f(x) q_f(x)\ \ \text{and} \ \
d_{f_1}(x) = d_{f_2}(x)/ (1+x)
\end{equation}
This derivation follows from the relationship between the original encoder and its dual form: $\frac{q_f(x)}{a_f(x)} = \frac{q_f(x)z_f(x)}{a_f(x)z_f(x)} = \frac{d_{f_2}(x)}{x^{N-1}+1}$.

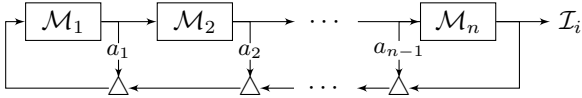
\begin{figure} 
     \centering
           \begin{tikzpicture}[auto, node distance=2cm, >=latex']

            \node [draw, rectangle, minimum height=0.5cm, minimum width=1cm, inner sep=0pt] (S1) at (0, 0) {$\mathcal{M}_1$};
            \node [draw, rectangle, minimum height=0.5cm, minimum width=1cm, right of=S1, node distance=1.75cm, inner sep=0pt] (S2) {$\mathcal{M}_2$};
            \node [inner sep=5pt, right of=S2, node distance=1.75cm] (dots) {$\cdots$};
            \node [draw, rectangle, minimum height=0.5cm, minimum width=1cm, right of=dots, node distance=1.75cm, inner sep=0pt] (Sn) {$\mathcal{M}_n$};


            \node [inner sep=-1pt, below right=0.5cm and 0.1cm of S1] (oplus1d) {$\triangle$};
            \node [inner sep=-1pt, below right=0.5cm and 0.1cm of S2] (oplus2d) {$\triangle$};
            \node [inner sep=-1pt, below left=0.5cm and 0.15cm of Sn] (oplusn-1d) {$\triangle$};
            \node [inner sep=5pt, right of=oplus2d, node distance=1cm] (dotsd) {$\cdots$};

           
            \node [ inner sep=0.5pt, above =0.25cm of oplus1d] (q1) {\small {$a_1$}};
            \node [ inner sep=0.5pt, above =0.25cm of oplus2d] (q2) {\small {$a_2$}};
            \node [ inner sep=0.5pt, above =0.25cm of oplusn-1d] (qn-1) {\small {$a_{n-1}$}};           
            
            \node [right of=Sn, node distance=1.5cm] (output) {$\mathcal{I}_i$};
            
            
            \draw [->] (S1) -- (S2);
            \draw [->] (S2) -- (dots);
            \draw [->] (dots) -- (Sn);

            \draw [-] (S1)   -|  (q1);
            \draw [->] (q1)   --  (oplus1d);
            \draw [-] (S2)   -|  (q2);
            \draw [->] (q2)   --  (oplus2d);
            \draw [-] (Sn)   -|  (qn-1);
            \draw [->] (qn-1)   --  (oplusn-1d);

            \draw [->] (Sn) -- ++(0.815,0)  |-  (oplusn-1d);
            \draw [->] (oplusn-1d)   --  (dotsd);
            \draw [->] (dotsd)   --  (oplus2d);
            \draw [->] (oplus2d)   --  (oplus1d);
            \draw [->] (oplus1d)  -- ++(-1.5,0)   |-  (S1);
            \draw [->] (Sn)  -- (output);

        \end{tikzpicture}    
    \caption{Recursive label synthesizer of the LMAP decoding for $\mathbf{g}(x)$}  
    \label{Fig::general::RLS}
    \vspace{-1.5em}
\end{figure}

The register labels $\mathcal{I}$ and $\mathcal{J}$ are determined through a Recursive Label Synthesizer (RLS)  shown in Fig \ref{Fig::general::RLS}. First, we introduce the Symmetric Difference Operation (SDO) $\triangle$, which retains only elements that differ between two sets (e.g., $\{1,3\} \triangle \{1,2\} = \{2,3\}$). RLS operates like a  rate-1 feedback encoder $\frac{1}{a_f(x)}$ but replaces multiplication with SDO operations. RLS initializes $\mathcal{I}' = \{n\}$, and $n$ registers as $\mathcal{M}_i = \{i\}$ for $i = 1,\ldots,n$. In each shift, $\mathcal{M}_n$ generates one RLS output. After $N-2$ shifts, the RLS outputs generate all necessary register labels $\mathcal{I}' = [\mathcal{I}_1',\mathcal{I}_2',\ldots,\mathcal{I}_{N-1}']$.

To determine $\mathcal{I}$ from $\mathcal{I}'$, we define set $\mathcal{U}_f \triangleq \{i : r_i \neq 0, 1 \leq i \leq n-1\}$, where $r_i = a_i\oplus q_i$. In fact, $\mathcal{U}_f$ indicates which encoder memory determine the encoding output. We then circularly shift $\mathcal{I}'$ to obtain $\mathcal{I}$ such that $\mathcal{U}_f$ appears as the last element, i.e.,
\begin{equation}  \small
\mathcal{I} = (\mathcal{I}'_{j+1}, \mathcal{I}'_{j+2}, \ldots, \mathcal{I}'_{N-1}, \mathcal{I}'_1, \mathcal{I}'_2, \ldots,\mathcal{I}'_j),
\end{equation}
satisfying $\mathcal{I}'_j = \mathcal{U}_f$. 
Once $\mathcal{I}$ is obtained, the sequence $\mathcal{J}$ is derived through $\mathcal{J}_i = \mathop{\bigtriangleup}\limits_{1\leq j \leq i } \mathcal{I}_j$ for $i = 1,2,\ldots,N-2$, following from the relationship $(1+x)d_{f_1}(x) = d_{f_2}(x)$.

Finally, the SUR label $\mathcal{S}$ is identified as the unique label that appears in $\mathcal{I}$ but not in $\mathcal{J}$, i.e., $\mathcal{S} = \mathcal{I} \setminus \mathcal{J}$. The coefficient $d_s$ of SUR is then determined based on whether $\mathcal{S}$ contains the first memory index: $d_s = 1$ if $1 \notin \mathcal{S}$ and $d_s = 0$ if $1 \in \mathcal{S}$.

\begin{example}
Consider the $(1,\frac{1+x+x^3}{1+x^2+x^3})$ code. RLS is initialized as $\mathcal{M}_i = \{i\}$ and $\mathcal{I}' = (\{3\})$. We obtain after 6 shifts of RLS:
$\mathcal{I}' = (\{1,2\}, \{2,3\},\{1,2,3\},\{1,3\},\{1\},\{2\}, \{3\})$. 

Then, since $\mathcal{U}_f = \{1,2\}$, we can have $\mathcal{I} =(\{2,3\},\{1,2,3\} ,\allowbreak \{1,3\},\{1\},\{2\}, \{3\}, \{1,2\})$. From $\mathcal{I}$, we compute 
$\mathcal{J} = (\{1,2\},\allowbreak \{1,3\}, \{2\}, \{1,2,3\},\{2,3\},\{3\})$. 
Finally, the SUR label is determined as $\mathcal{S} = \mathcal{I} \setminus \mathcal{J} = \{1\}$.
\end{example}

\vspace{-0.5em}
\subsection{LMAP Decoding Process}
\vspace{-0.3em}
The LMAP decoder processes inputs $(\hat{\mathbf{x}}_{\mathbf{c}_1},\ldots,\hat{\mathbf{x}}_{\mathbf{c}_L})$ in the forward direction and $(\hat{\mathbf{x}}_{\mathbf{c}_L},\ldots,\hat{\mathbf{x}}_{\mathbf{c}_1})$ in the backward direction. All registers $F_{\mathcal{J}_i}^{(1)}[k]$, $F_{\mathcal{I}_j}^{(2)}[k]$, $B_{\mathcal{J}_i}^{(1)}[k]$, and $B_{\mathcal{I}_j}^{(2)}[k]$ are initialized to 1. Following similar principles as illustrated in Section \ref{sec::7O5}, the decoding process consists of two stages:

\subsubsection{Register Updates}
 The forward decoders perform the following updates sequentially. $\mathrm{DF}_1$ updates registers as
$$\small F_{\mathcal{J}_{i+1}}^{(1)}[k+1] = F_{\mathcal{J}_{i}}^{(1)}[k]\cdot \hat{x}_{c_k^1} \cdot (\hat{x}_{c_k^2})^{d_i^{(1)}}.$$
Next, $\mathrm{DF}_2$ updates registers according to
$$\small F_{\mathcal{I}_{j+1}}^{(2)}[k+1] = F_{\mathcal{I}_{j}}^{(2)}[k]\cdot (\hat{x}_{c_k^2})^{d_i^{(2)}}.$$
Then, for each pair of identical labels $\mathcal{I}_j \in \mathcal{I}$ and $\mathcal{J}_i \in \mathcal{J}$, where $\mathcal{I}_j=\mathcal{J}_i$, compute combined values:
$$\small F_{i,j}^{\mathrm{c}}[k+1] = \frac{1}{\lambda_k}(F_{\mathcal{J}_i}^{(1)}[k+1]+ F_{\mathcal{I}_j}^{(2)}[k+1])$$
where $\lambda_k =1+ F_{\mathcal{J}1}^{(1)}[k] \cdot \hat{x}_{c_k^1} \cdot \hat{x}_{c_k^2}$. Update both decoders with combined values, i.e., 
$F_{\mathcal{J}_i}^{(1)}[k+1] = F_{\mathcal{I}_j}^{(2)}[k+1] = F_{i,j}^{\mathrm{c}}[k+1]$.

In particular, let $\mathcal{I}_s$ be the identical counterpart of SUR label $\mathcal{S}$ in $\mathcal{I}$. Similar combining and updating applies to the SUR, i.e., $F_{s}^{\mathrm{c}}[k+1] = \frac{1}{\lambda_k}(F_{\mathcal{S}}^{(1)}[k+1]+ F_{\mathcal{I}_s}^{(2)}[k+1]),$
and $F_{\mathcal{S}}^{(1)}[k+1] = F_{\mathcal{I}_s}^{(2)}[k+1] = F_{s}^{\mathrm{c}}[k+1].$

The backward decoders $\mathrm{DB}1$ and $\mathrm{DB}2$ follow similar update rules but operate in reverse time order, processing from time $k = L-1$ down to $k = 1$. The normalization factor for backward decoding is $\rho_k = 1 + B_{\mathcal{J}_{N-2}}^{(1)}[k+1] \cdot \hat{x}_{c_k^1} \cdot \hat{x}_{c_k^2}$.

\subsubsection{Output Generation}

After updating all register values, the final bidirectional LLR of $b_k$ is obtained as in \eqref{equ::biLLR}, with $\mu_k$ and $\delta_k$ particularly given by
$$ \small
    \mu_k \!=\! \sum_{i = 0}^{N-2} \!\left(\!\hat{x}_{c_k^2}\!\right)\!^{d_i^{(1)}} \!\! F_{\mathcal{J}_i}^{(1)}[k] B_{\mathcal{J}_{i+1}}^{(1)}[k] + (\hat{x}_{c_k^2})^{d_s} F_{\mathcal{S}}^{(1)}[k]   B_{\mathcal{S}}^{(1)}[k] ,
$$
$$  \small
    \delta_k \!=\! 1 \!+\! \sum_{j = 1}^{N-2} \!\left(\!\hat{x}_{c_k^2}\!\right)\!^{d_{N\!-\!1\!-\!j}^{(2)}}  F_{\mathcal{I}_j}^{(2)}[k] B_{\mathcal{I}_{j+1}}^{(2)}[k] + F_{\mathcal{I}_{N-1}}^{(1)}[k] B_{\mathcal{I}_{1}}^{(1)}[k].
$$
For notational convenience, we introduced auxiliary variables $F_{\mathcal{J}_0}^{(1)}[k] = B_{\mathcal{J}_{N-1}}^{(1)}[k] = 1$ and $d_0^{(1)} = d_{N-2}^{(1)} = 1$ in computing $\mu_k$ and $\delta_k$ above, which are not used in the actual decoder structure.  The procedure of LMAP is outlined in Algorithm \ref{algo::LMAP}.

\begin{remark}
Terms $(\hat{x}_{c_k^2})^{d_i^{(1)}}$ and $(\hat{x}_{c_k^2})^{d_j^{(2)}}$ do not increase computational complexity, as they represent fixed connection patterns pre-determined by decoder polynomials.
\end{remark}

\begin{algorithm}[t]
\small
\caption{LMAP Decoding}
\label{algo::LMAP}
\begin{algorithmic}[1]
\REQUIRE Input $(\hat{\mathbf{x}}_{\mathbf{c}_1},\ldots,\hat{\mathbf{x}}_{\mathbf{c}_L})$, decoder parameters $\mathcal{D}$
\ENSURE Decoded LLRs $(L_{b_1},\ldots,L_{b_L})$
\STATE Initialize all registers to 1
\FOR{$k = 1$ to $L$}
\STATE Update forward registers $F_{\mathcal{J}_i}^{(1)}[k], F_{\mathcal{I}_j}^{(2)}[k]$
\ENDFOR
\FOR{$k = L$ to $1$}
\STATE Update backward registers $B_{\mathcal{J}_i}^{(1)}[k], B_{\mathcal{I}_j}^{(2)}[k]$
\STATE Compute $\delta_k$ and $\mu_k$, and then compute LLR $L_{b_k}$.
\ENDFOR
\end{algorithmic}
\end{algorithm}

\vspace{-1em}
\section{Complexity Analysis}
\vspace{-0.3em}
The LMAP decoder achieves significant complexity reduction compared to the traditional BCJR algorithm. For computational complexity per time step, LMAP requires only $4N+25$ additions and $13N+17$ multiplications, compared to BCJR's $24N+15$ additions and $26N+17$ multiplications. 

The storage requirements are also more efficient. Let $n_q$ denote the number of quantization bits for soft information. While BCJR requires $L(4n_qN + n_q) + N(6+2m)$ memory units with random access patterns to metric tables, LMAP needs only $L(2n_qN + n_q) + N(m+4n_q+2) - 4(n_q + 1)$ units with sequential register access. 
Moreover, LMAP's shift-register structure allows direct hardware mapping and high cache efficiency, requiring only $6(N-1)$ sequential register accesses per time step, compared to BCJR's $12N$ random metric accesses random path lookups per step. A comprehensive comparison among these decoders is summarized in Table \ref{tab:complexity}.

\begin{table}
\footnotesize	
\tabcolsep=0.11cm
\centering
\caption{Computational and Storage Complexity Comparison}

\begin{tabular}{|l|c|c|}
\hline
& \textbf{Operations/Step} & \textbf{Memory Access/Step} \\
\hline
BCJR & \begin{tabular}{@{}c@{}} $24N+15$ add \\ $26N+17$ mult \end{tabular} & $12N$ random \\
\hline
LMAP & \begin{tabular}{@{}c@{}} $4N+25$ add \\ $13N+17$ mult \end{tabular} & $6(N-1)$ sequential \\
\hline
\end{tabular}
\label{tab:complexity}
\vspace{-1em}
\end{table}

\vspace{-1.2em}
\section{Simulation Results}
\vspace{-0.5em}

The performance and computational efficiency of the proposed LMAP algorithm are evaluated through extensive simulations in MATLAB 2021a, collecting at least 3000 bit errors per data point to ensure statistical reliability. We examine the algorithm's effectiveness across recursive systematic codes (RSC) and non-systematic codes (NSC) with tail-biting.




\subsubsection{Recursive Systematic Codes}
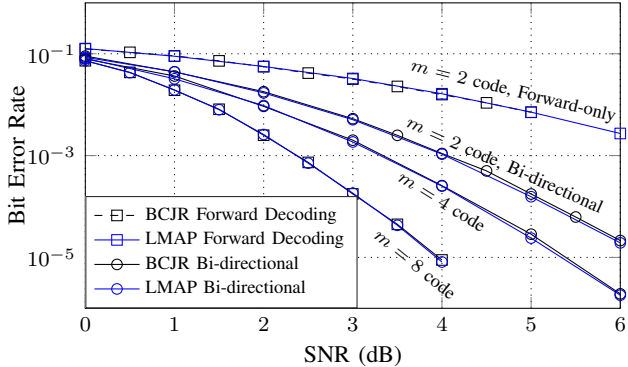
\begin{figure}   [t]
     \centering
        \begin{tikzpicture}
    \definecolor{applegreen}{rgb}{0.55, 0.71, 0.0}
    \definecolor{english}{rgb}{0.0, 0.5, 0.0}
    \begin{axis}[%
    width=2.8in,
    height=1.6in,
    at={(0.822in,0.529in)},
    scale only axis,
    xmin=0,
    xmax=6,
    xlabel style={at={(0.5,1ex)},font=\color{white!15!black},font = \small},
    xlabel={SNR (dB)},
    ymode=log,
    ymin=1e-06,
    ymax=1,
    ylabel style={at={(2ex,0.5)},font=\color{white!15!black},font = \small},
    ylabel={Bit Error Rate},
    axis background/.style={fill=white},
    tick label style={font=\footnotesize},
    xmajorgrids,
    ymajorgrids,
    yminorgrids,
    major grid style={dotted,black},
    minor grid style={dotted},
    legend style={at={(0.0,0.0)}, anchor=south west, legend cell align=left, align=left, draw=white!15!black,font = \scriptsize	,row sep=-1pt, legend columns=1}
    ]

    \addplot [color=black, dashed, mark=square, mark options={solid, black}]
      table[row sep=crcr]{%
    0	0.124355158730159\\
    0.5	0.106650204359673\\
    1	0.0907953886310905\\
    1.5	0.0724971064814815\\
    2	0.0558131241084165\\
    2.5	0.0417110874200426\\
    3	0.03185389544345\\
    3.5	0.0228081217162872\\
    4	0.0158115388506677\\
    4.5	0.0108718341219037\\
    5	0.00708552512243787\\
    };
    \addlegendentry{BCJR Forward Decoding} 
    
    \addplot [color=blue, mark=square, mark options={solid, blue}]
      table[row sep=crcr]{%
    0	0.126535952669903\\
    1	0.0895061176302232\\
    2	0.056279820849838\\
    3	0.0327019673503558\\
    4	0.0163981333298352\\
    5	0.0071383502352565\\
    6	0.00271843139983994\\
    };
    \addlegendentry{LMAP Forward Decoding}
    
    \addplot [color=black, mark=o, mark options={solid, black}]
      table[row sep=crcr]{%
    0	0.09\\
    1	0.044\\
    2	0.018\\
    3	0.0053\\
    3.5	0.0025\\
    4	0.0011\\
    4.5	0.0005\\
    5	0.000176\\
    5.5	6.216e-05\\
    6	2.143e-05\\
    6.5	5.83e-06\\
    };
    \addlegendentry{BCJR Bi-directional }

    \addplot [color=blue, mark=o, mark options={solid, blue}]
      table[row sep=crcr]{%
    0	0.0852893013100437\\
    1	0.0443565550510783\\
    2	0.0170389500654165\\
    3	0.00506617122665975\\
    4	0.00106836876912986\\
    5	0.000155461499275674\\
    6	1.93046277853547e-05\\
    };
    \addlegendentry{LMAP Bi-directional }

    \addplot [color=black, mark=o, mark options={solid, black}]
      table[row sep=crcr]{%
    0	0.0788454610951009\\
    1	0.0365975935828877\\
    2	0.0092627879403794\\
    3	0.0020076174743025\\
    4	0.000256995926823768\\
    5	2.85848615685925e-05\\
    6	1.92356781234664e-06\\
    };
    
    \addplot [color=blue, mark=o, mark options={solid, blue}]
      table[row sep=crcr]{%
    0	0.0816243489583333\\
    1	0.0322088653212521\\
    2	0.00960947109471095\\
    3	0.00185099114247821\\
    4	0.000251830985461669\\
    5	2.39375e-05\\
    6	1.8125e-06\\
    7	7.03125e-08\\
    };

    \addplot [color=black, mark=square, mark options={solid, black}]
    table[row sep=crcr]{%
    -1	0.142987654321098\\
    -0.5	0.107123456789012\\
    0	0.0728765432109876\\
    0.5	0.0431234567890123\\
    1	0.0192345678901234\\
    1.5	0.00812345678901234\\
    2	0.00251234567890123\\
    2.5	0.00073456789012345\\
    3	0.000176543209876543\\
    3.5	4.4567890e-05\\
    4	8.901234e-06\\
    };

    \addplot [color=blue, mark=o, mark options={solid, blue}]
      table[row sep=crcr]{%
    -1	0.143623477616853\\
    -0.5	0.106316896024465\\
    0	0.0731907894736842\\
    0.5	0.0425908221797323\\
    1	0.0195487892880051\\
    1.5	0.00797472094950283\\
    2	0.00259481939991636\\
    2.5	0.00070941798261536\\
    3	0.000182987625009984\\
    3.5	4.2734375e-05\\
    4	8.125e-06\\
    };
    
    \node[rotate=-39] at (axis cs: 3.7,7.5e-6) {\scriptsize $m=8$ code};

    \node[rotate=-32] at (axis cs: 4,1.3e-4) {\scriptsize $m=4$ code};

    \node[rotate=-24] at (axis cs: 4.7,6.5e-4) {\scriptsize $m=2$ code, Bi-directional};

    \node[rotate=-14] at (axis cs: 4.8,1.8e-2) {\scriptsize $m=2$ code, Forward-only};
 
    \end{axis}    
    
    \end{tikzpicture}%
	\vspace{-0.5em}
    \caption{BER performance comparison for decoding RSC.}
    \vspace{-1.5em}
	\label{Fig::Sim::RSC}  
\end{figure}

We examine three RSC codes with increasing complexity: $(1,\frac{7}{5})_{\mathrm{oct}}$ ($m=2$), $(1,\frac{23}{25})_{\mathrm{oct}}$ ($m=4$), and $(1,\frac{561}{573})_{\mathrm{oct}}$ ($m=8$). Both forward-only and bi-directional variants are evaluated for the $(1,\frac{7}{5}){\mathrm{oct}}$ code. For $m=4$ and $m=8$, we only plot the bi-directional decoding for a clear illustration. As shown in Fig. \ref{Fig::Sim::RSC}, the bi-directional LMAP decoder achieves identical performance to the BCJR algorithm across all code configurations. This demonstrates the optimality of the LMAP decoder and its equivalence to the BCJR MAP decoding.

\subsubsection{Non-Systematic Codes}

We evaluate LMAP with a NSC code and implement the tail-biting configurations. For tail-biting codes with message length $L$, we set the initial state $s_0$ to the last $m+1$ (constraint length) bits of the message in the encoding, ensuring the encoder returns to state $s_0$ after encoding, thus creating a circular structure. The LMAP decoder handles this circularity through multiple decoding passes. Specially, the decoder initializes all registers to 0 (rather than 1 originally). Then, it processes the received sequence cyclically (positions 1 to $L$, then $L+1$ to $2L$, etc.) After five iterations, it computes final LLRs using register values from the last $L$ time steps. Note that for NSC, while the decoder polynomials and register labels in $\mathcal{I}$ and $\mathcal{J}$ are generated following the same way as RSC, an additional register label mapping is required when computing $\delta_k$ and $\mu_k$ for the bi-directional LLR \eqref{equ::biLLR}. As both codeword bits are now coded (non-systematic), the $\hat{x}_{c_k^1}$ terms in \eqref{equ::biLLR} are also replaced with appropriate memory state combinations. The details of these modifications are omitted here due to space limitation.

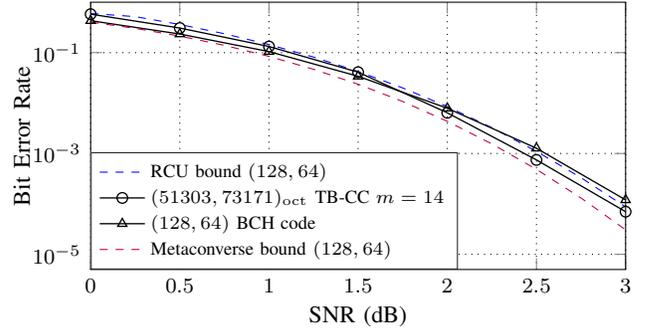
\begin{figure} [t]
     \centering
        \begin{tikzpicture}
    \definecolor{applegreen}{rgb}{0.55, 0.71, 0.0}
    \definecolor{english}{rgb}{0.0, 0.5, 0.0}
    \begin{axis}[%
    width=2.8in,
    height=1.4in,
    at={(0.822in,0.529in)},
    scale only axis,
    xmin=0,
    xmax=3,
    xlabel style={at={(0.5,1ex)},font=\color{white!15!black},font = \small},
    xlabel={SNR (dB)},
    ymode=log,
    ymin=0.5e-5,
    ymax=1,
    ylabel style={at={(2ex,0.5)},font=\color{white!15!black},font = \small},
    ylabel={Bit Error Rate},
    axis background/.style={fill=white},
    tick label style={font=\footnotesize},
    xmajorgrids,
    ymajorgrids,
    yminorgrids,
    major grid style={dotted,black},
    minor grid style={dotted},
    legend style={at={(0,0)}, anchor=south west, legend cell align=left, align=left, draw=white!15!black,font = \scriptsize	,row sep=-1pt, legend columns=1}
    ]

    \addplot [color=blue, dashed]
      table[row sep=crcr]{%
    0	0.598051442160651\\
    0.25	0.517561391100383\\
    0.5	0.361691910793777\\
    0.75	0.237100933354601\\
    1	0.144772286129236\\
    1.25	0.0817304803690512\\
    1.5	0.0423356503170512\\
    1.75	0.019964460204284\\
    2	0.00850383099651646\\
    2.25	0.0032463178995015\\
    2.5	0.00110235295182097\\
    2.75	0.00033067539636614\\
    3	8.71215624409524e-05\\
    };
    \addlegendentry{RCU bound  $(128,64)$}


    \addplot [color=black, line width=0.5pt, mark=o, mark options={solid, black}]
      table[row sep=crcr]{%
    0	0.579710144927536\\
    0.5	0.307692307692308\\
    1	0.132362673726009\\
    1.5	0.0409416581371546\\
    2	0.0064059447166971\\
    2.5	0.000745269891145775\\
    3	7e-05\\
    };
    \addlegendentry{$(51303,73171)_{\mathrm{oct}}$ TB-CC $m = 14$}

    \addplot [color=black, line width=0.5pt, mark=triangle, mark options={solid, black}]
      table[row sep=crcr]{%
    0	0.43\\
    0.5	0.234\\
    1	0.105\\
    1.5	0.034\\
    2	0.0079\\
    2.5	0.00127\\
    3	0.00012\\
    };
    \addlegendentry{$(128,64)$ BCH code}

    \addplot [color=purple, dashed]
      table[row sep=crcr]{%
    0	0.40264455028142\\
    0.25	0.302779050241713\\
    0.5	0.212663259367559\\
    0.75	0.139531532469658\\
    1	0.0844500678530219\\
    1.25	0.0467893545302211\\
    1.5	0.0236573960180887\\
    1.75	0.0107416764961012\\
    2	0.00434838098695346\\
    2.25	0.0015463097135208\\
    2.5	0.000484409929882316\\
    2.75	0.000130899949900953\\
    3	3.05981036835679e-05\\
    };
    \addlegendentry{Metaconverse bound $(128,64)$}

    \draw[draw=black, thick] (axis cs: 30,4) rectangle ++(40,20);
    \draw[->,black, dashed, thick] (axis cs: 34,6) -- (axis cs: 47,6);
 
    \end{axis}    
    
    \end{tikzpicture}%
        \vspace{-0.5em}
    \caption{Tail-biting NSC against PPV bounds with $L = 64$.}
    \vspace{-1em}
	\label{Fig::Bounds::128}  
\end{figure}

We evaluate performance of LMAP for tail-biting codes against finite blocklength bounds \cite{erseghe2016coding} with block length $128$ and message length $L=64$. As shown in Fig. \ref{Fig::Bounds::128}, TB-CC $(51303,73171){\mathrm{oct}}$ with memory $m=14$ is examined. This TB-CC demonstrates strong error-correction capabilities, with block error rate (BLER) outperforming the classical $(128,64)$ BCH code at higher SNRs. These results show that well-designed TB-CCs with LMAP decoding can achieve excellent performance for short blocklength applications.


 \begin{figure}   [t]
     \centering
        \begin{tikzpicture}
    \definecolor{applegreen}{rgb}{0.55, 0.71, 0.0}
    \definecolor{english}{rgb}{0.0, 0.5, 0.0}
    \begin{axis}[%
    width=2.8in,
    height=1.4in,
    at={(0.822in,0.529in)},
    scale only axis,
    xmin=1,
    xmax=12,
    xlabel style={at={(0.5,1ex)},font=\color{white!15!black},font = \small},
    xlabel={Momory length $m$},
    ymode=log,
    ymin=1e-1,
    ymax=1e5,
    ylabel style={at={(2ex,0.5)},font=\color{white!15!black},font = \small},
    ylabel={Decoding time (ms)},
    axis background/.style={fill=white},
    tick label style={font=\footnotesize},
    xmajorgrids,
    ymajorgrids,
    yminorgrids,
    major grid style={dotted,black},
    minor grid style={dotted},
    legend style={at={(0.05,0.95)}, anchor=north west, legend cell align=left, align=left, draw=white!15!black,font = \scriptsize	,row sep=-1pt, legend columns=1}
    ]

    \addplot [color=black, line width=0.5pt, mark=square, mark options={solid, black}]
      table[row sep=crcr]{%
    2	8.166\\
    3	16.133\\
    4	32.3\\
    5	62.289\\
    6	127.097\\
    7	255.68\\
    8	523.183\\
    9	1050.692\\
    10	2096\\
    11	4272.65\\
    };
    \addlegendentry{BCJR Bi-directional}

    \addplot [color=black, line width=0.5pt, mark=square, mark options={solid, black}]
      table[row sep=crcr]{%
    2	0.678\\
    3	0.706\\
    4	0.751\\
    5	0.857\\
    6	0.965\\
    7	1.178\\
    8	1.583\\
    9	2.348\\
    10	5.5\\
    11	12.649\\
    };
    \addlegendentry{LMAP Bi-directional}

    \draw[draw=black, thick] (axis cs: 30,4) rectangle ++(40,20);
    \draw[->,black, dashed, thick] (axis cs: 34,6) -- (axis cs: 47,6);
 
    \end{axis}    
    
    \end{tikzpicture}%
	\vspace{-0.5em}
    \caption{Decoding time comparison between LMAP and BCJR with respect to code memory order of RSCs.}
    \vspace{-1.5em}
	\label{Fig::Time::Memory}  
\end{figure}
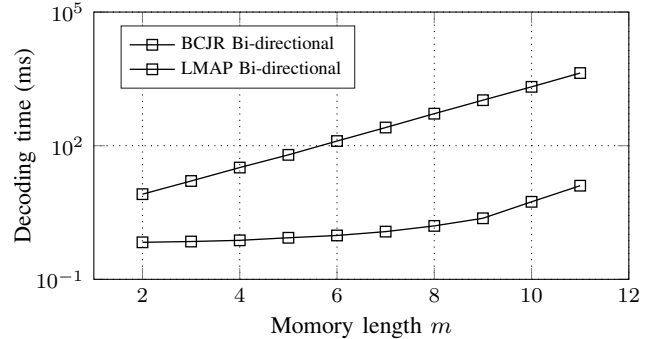

\subsubsection{Decoding Efficiency}
LMAP achieves significant computational speedup compared to BCJR. For RSC codes, the speedup scales with code complexity, which are 12$\times$ for the basic $(1,\frac{7}{5})_{\mathrm{oct}}$ code and 331$\times$ for the $(1,\frac{561}{573})_{\mathrm{oct}}$ code. Similar efficiency gains are observed for NSC; for instance, there is a 132$\times$ speedup for the $(171,133)_{\mathrm{oct}}$ code.

\begin{table}
\vspace{-1.5em}
\footnotesize	
    	\tabcolsep=0.11cm
\centering
\caption{Decoding Time Comparison}
\vspace{-0.75em}
\begin{tabular}{|l|c|c|c|}
\hline
\textbf{Code} & \textbf{LMAP (ms)} & \textbf{BCJR (ms)} & \textbf{Ratio} \\
\hline
$(1,\frac{7}{5}){\mathrm{oct}}$ RSC & 0.67 & 8.16 & 12× \\
\hline
$(1,\frac{23}{25}){\mathrm{oct}}$ RSC & 0.75 & 32.30 & 43× \\
\hline
$(1,\frac{561}{573}){\mathrm{oct}}$ RSC & 1.58 & 523.18 & 331× \\
\hline
            $(171,133)_{\mathrm{oct}}$ NSC & 0.96 & 127.10 & 132× \\
\hline
$(5621,7173)$ TB-CC & 20.12 & 2231.75 & 110× \\
\hline
\end{tabular}
\label{table:decoding-time}
\vspace{-1.5em}
\end{table}

As shown in Fig. \ref{Fig::Time::Memory}, while computation time of BCJR grows exponentially with memory order, LMAP maintains near-linear scaling up to order 9, and beyond that exhibits exponential growth but with a much lower slope. At memory order 11, for instance, LMAP requires only 12.6 ms compared to BCJR's 4272.6 ms. Table \ref{table:decoding-time} summarizes efficiency gains for several codes.

\vspace{-1.2em}
\section{Conclusion}
\vspace{-0.3em}
 In this paper, we developed a linear representation of BCJR MAP algorithm, referred to as the linear MAP decoding (LMAP) for a rate 1/2 binary convolutional code. LMAP decoder consists of the MAP forward and backward decoding, each comprising two constituent dual encoders, associated with two constituent codes. The bidrectional MAP decoding output can be obtained by combining the contents of respective forward and backward dual encoders. With the linear presentation of the decoding structure using shift registers, LMAP significantly reduced decoding delay and at the same time achieves the same performance as the BCJR MAP algorithm. It unlocks the decoding of very large state codes (VLSC) with near-capacity approach performance. 

\ifCLASSOPTIONcaptionsoff
\newpage
\fi

\bibliographystyle{IEEEtran}
\bibliography{reference}

\end{document}